# Microcontroller based automated life savior - Medisûr

Soumick Chatterjee, Pramod George Jose, Priyanka Basak, Ambreen Athar, Bindhya Aravind, Romit S. Beed & Rana Biswas
*Dept. of Computer Science, St. Xavier's College (Autonomous), Kolkata*

ABSTRACT: With the course of progress in the field of medicine, most of the patients' lives can be saved. Only thing required is the proper attention in proper time. Our wearable solution tries to solve this issue by taking the patients vitals and transmitting them to the server for live monitoring using mobile app along with patient's current location. In case of an emergency, that is if any vitals show any abnormalities, a SMS is sent to the caregiver of the patient with the patient's location so that he can reach there on time.

## 1 INTRODUCTION

### 1.1 Objective

The 21st century has witnessed a major paradigm shift in diagnosis technology in the field of medicine. Embedded technology has been effectively used for successful diagnosis, leading to more accurate treatments. High speed calculations, more complex testing methodologies, more accurate and reliable results have forever enhanced diagnosis process.

Today's world is becoming increasingly dependent on heuristic data; medical treatment is no exception. Large volume of data is generated about patients, symptoms and diagnosis. In the previous decade, doctors would have recorded certain basic information about the patients in a prescription. In contrast, these days he will have a long stream of digital data, from minute details about the patient to high-definition medical images.

Since the world is going green, this data is needs to be made available in digital format. This in turn has broadened the application of informatics to improve health care and has also contributed in medical research. Today, informatics is being applied at every stage of health care from basic research to care delivery and includes many specializations such as bioinformatics, medical informatics, and biomedical informatics, and is often referred to as "in silico" research. Informatics specializes on introducing better means of using technology to process information.

The medical field has developed sufficiently to ensure critical patients in majority of the cases, can survive cardiac arrest or similar issues. The person can return to normalcy, with proper diagnosis and medication. But the treatment has to be given at the right time, which is during the initial stage of an attack. This becomes difficult to achieve when the person is alone, or when the patient is a senior citizen. Medisûr tries solves the problem by being an automated companion who watches the patient's body vitals and calls for help when in danger.

The Medisûr gadget monitors the body vitals and updates the data in a database. Caretaker of the patient can live monitor this information. In case, if the vitals are in the danger zone, an alert message is sent to the caretaker. So, the patient receives proper treatment on time.

### 1.2 Working Principal

The electronic gadget under consideration consists of a wearable device and a receiving kit. The wearable device is supposed to detect and get the body vitals of the person wearing it, post which it sends the same to the receiving kit. The wearable band has sensors to serve this purpose and an emergency button in case of emergency. The receiving kit has three tier functionality. The first to track the current location of the person wearing it, second, it sends the collected information to a centralized server for live monitoring and also for future reference and lastly, to check whether the received vitals are within the normal range or not. If not, then it sends a SMS to the pre-configured number with the current location along with the current vitals for immediate assistance. Another scenario could be, if the patient feels any discomfiture, feels dizzy that can never be detected by any sensor. In that case, the patient presses the emergency button, then also a similar SMS will be sent to the pre fed mobile number stating the person requires attention. This being an embedded system, it has related software modules. There is a central server, where the current vitals & location are sent from the receiving kit. Server also stores daily data for future reference, from which a caregiver can get an idea about how the patient's health was in past 6 months. Again, there is a Multi-platform mobile App. The app fetches the latest vitals and location from the server and shows it on screen, so that caregiver can check anytime how his loved one is doing and where he is currently.

## 2 COMPONENTS

### 2.1 *ATmega328*

ATmega328 is an 8-bit AVR RISC-based microcontroller by Atmel which is used in the wearable gadget, to collect sensor information, then encrypt it and send it to the receiving kit.

### 2.2 *Arduino Uno*

Arduino Uno is a versatile microcontroller board based on the ATmega328 which is used in the receiving kit to receive sensor data and location from the wearable gadget and then it will send the data to the server. If required, this will also help in sending SMS. This can be replaced with just another ATmega328 microcontroller instead of this complete board.

### 2.3 *LM35 Temperature Sensor*

LM35 series are precision integrated-circuit temperature devices which gives an output voltage linearly proportional to the Centigrade temperature it receives. This LM35 Sensor is used to detect body temperature in the prototype.

### 2.4 *M212 Pulse Sensor*

Pulse Sensor which is used in this implementation is a photoplethysmograph, which is a well-known medical device used for non-invasive heart rate monitoring. Sometimes, it measures blood-oxygen levels (SpO2), sometimes they don't. The heart pulse signal that comes out of a photoplethysmograph is an analog fluctuation in voltage, and it has a predictable wave shape which is shown in Figure 1.

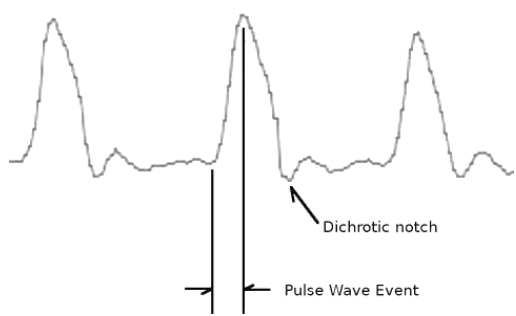

Figure 1. Pulse Wave

The depiction of the pulse wave is called a photoplethysmogram, or PPG. The goal is to find successive moments of instantaneous heart beat and measure the time between them, called the Inter Beat Interval (IBI), which can be archived by following the predictable shape and pattern of the PPG wave. In this implementation, it is used to measure the pulse in the wearable gadget and send it to the ATmega328 for further processing.

### 2.5 *434MHz Radio Frequency Receiver (Rx) & Transmitter (Tx)*

RF transmitter and receiver module is used to transmit and receive the radio frequency for the wireless communication which is used to send sensor information from the wearable gadget to the receiving kit.

### 2.6 *NEO6MV2 GPS Module*

The NEO-6 module series is a family of GPS receivers which consist of high performance u-blox 6 positioning engine, giving it receivers excellent navigation performance even in the most challenging environments. This is used to triangulate location of the user and send it to the Arduino for further processing as required.

### 2.7 *SIM300 GSM Module*

A GSM Modem is a device that modulates and demodulates the GSM signals and in this particular case 2G signals. The modem which is used in this implementation is a SIMCOM SIM300 which is a Tri-band GSM/GPRS Modem. The TTL interface present in it allows to be directly interfaced with the Arduino present in the receiving kit, using which, the device is connecting to the designated web service and sending sensor plus location information. Also this GSM Module is used to send SMS when required.

### 2.8 *AES 256bit Encryption*

The Advanced Encryption Standard (AES), also known as Rijndael, is a specification for the encryption of electronic data. In this implementation AES with 256 bits Key Size is used for encrypting data while sending from wearable gadget to the receiving kit and from receiving kit to the server.

### 2.9 *Microsoft Azure Virtual Machine*

Microsoft Azure is a cloud computing platform and infrastructure created by Microsoft for building, deploying, and managing applications and services through a global network of Microsoft-managed data centers. This is used to host the Web Service and al-so to host Microsoft SQL Server in which the data received from receiving kit is stored for further use. Mobile App can access this to show the live information and also can be used to track patient history.

## 3 WORKING MECHANISM

### 3.1 Module 1: Wearable Gadget

First module is a wearable gadget in the form of a wrist band. This wearable gadget consists of many bio sensors. For this implementation, a LM-35 is used as the temperature sensor. For sensing the pulse, M212 heart rate sensor is used. Both the sensors are placed on the inner side of the band, placed strategically so that it directly touches the skin. Pulse sensor should be placed in such a way that it can detect the pulse. Sensors are connected to a circuit board, where ATmega328 Microcontroller is placed. Data collected by the sensors are periodically sent to the microcontroller. A string is constructed by concatenating the sensor values. If the user feels unwell, that can't be detected by any sensor. For that, there is an Emergency Button. If the button is pressed then the value of the switch is one, or else zero. This value is also concatenated with the sensor values. For example, if the switched is pressed along with that, current body temperature is $98^O F$ and the current pulse rate is 77 bpm, then the string will be constructed as - 1,98,77 This string is now transmitted from the wrist band to the second module, which is the receiving kit, explained later. In this implementation, this transmission is done using 434Mhz RF-Transmitter. Here a problem may creep up, if two or more radio device with the same frequency is used in close range, it may interfere with each other. To overcome this, a unique prefix is added to the above string. In the prototype implementation, the prefix – "SXCMS" have been added. So, the string becomes – SXCMS:1,98,77 But this isn't enough. As wireless technology is used here, security is a concern. So, the above string is encrypted using AES 256bit encryption before transmitting using the unique device ID as the key. For the prototype, device ID is MS001. So, the final encrypted string that is to be transmitted from the wearable gadget becomes: - Ql05Ng7kM1M9PPY36BO/LbiLyJsVpDSp4hU4tVz9Nw= .

### 3.2 Module 2: Receiving Kit

Second module of this implementation is the receiving kit. In the receiving kit a 434Mhz RF-Receiver is connected to the Arduino, which receives the encrypted string from the wrist band and sends it to the Arduino Uno microcontroller board, which first decrypts it using the same AES scheme with the device ID as the key. This device ID is unique for each Wrist Band & Receiving Kit combo.

So, the receiving kit has the same key which the wrist band used to encrypt the string. So, in this prototype MS001 is used for decryption and the plain text SXCMS: 1,98,77 is obtained. First, the string is validated to check whether it is coming from the correct device. If the string is starting with SXCMS then it is understood that it is coming from the correct device. After this, the string is broken to extract the parameter values of switch, hear rate and temperature. In this implementation, NEO6MV2 GPS module is used for triangulating the location of the device. Utmost priority is given to the switch. If the switch value is 1, irrespective of the sensor values, a SMS is sent using the SIM300 module present, to the care giver stating that person wearing this device needs his attention along with the exact location and sensor information. Apart from this, using the same SIM300 module, a web service is called, which is hosted in the Microsoft Azure server, by sending a simple HTTP Get Request. While sending the request, the device ID is also sent to identify the patient uniquely. So, the plain text that is going to be sent via the get request is MS001,98,77,Park Street where MS001 is the device ID, $98^O F$ is the temperature, 77 is the pulse rate and Park Street is the triangulated location. For security reasons, encrypting this string is required and this is done using AES 256bit encryption with a global encryption key, in this case MedS, by which the cipher text is obtained as
vLbOEhcctSgLS8W66U6M6QdRnL52kIjcDN9ONNuHoWI= which is sent via the query string parameter. This is sent to the server at a regular interval, say every minute. Besides sending this information to the server, it is also checked if there is any anomalies in the pulse rate or temperature. Say, if the temperature goes beyond $99^O F$ or pulse rate goes below 40 bpm, then an SOS message is sent to the care giver with the current sensor values plus the triangulated location stating that the person wearing this device may require medical help immediately.

### 3.3 Software Modules

An embedded system is incomplete without software modules First soft-module is the web service which receives the cipher text from each of the Medisûr device, and decrypts it using the global key MedS. From the decrypted text, the device is first identified from the device ID. Consequently the individual sensor data are extracted, and then lastly the location.

Furthermore, this information is updated in the database, in the currentPatientData table. In every six hours, this information is stored in patientHistry table as well. Both web service and database are stored in the Microsoft Azure virtual machine, so as to access it centrally.

The last and one of the most vital part of our solution pertains to a cross platform mobile app. In the mobile app, a caregiver can add multiple persons whom he/she is looking after. When a person is selected, mobile app calls a WCF service, which is hosted in the same server. This provides latest data from the currentPatientData table. So, a caregiver can check the live vitals and location of patients, updated every minute. Using the app, the history of past 6 months can be checked, so that he/she can monitor his patient more efficiently.

## 4 RESULTS AND DISCUSSION

The Authors have implemented this concept by creating a minimalistic prototype consisting most of the components except the GPS Module, which is shown in Figure 2.

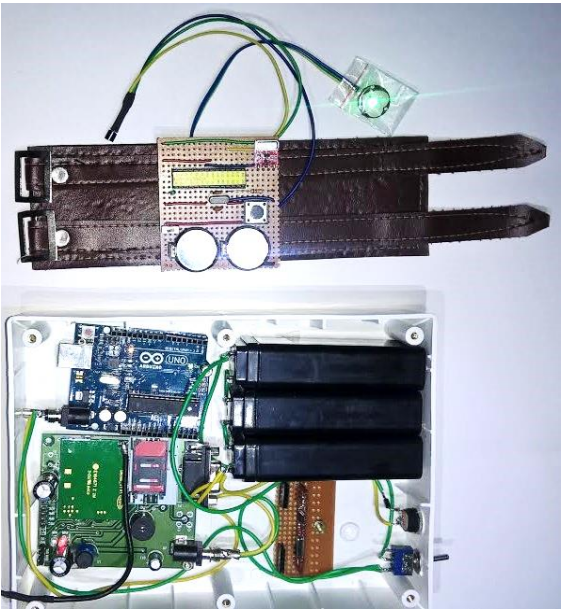

Figure 2. First Working Prototype

The outputs observed were close to the desired ones, but can be fine-tuned further. It was detecting heart-beat accurately, for body temperature, other sensors would give better results.

This prototype is quite bulky, and not very easy to carry. But this can be made quite compact by using latest industry based fabrication techniques.

For more accurate results, some of the components can be replaced with some superior ones. Like custom made body temperature sensor based on DS1624, DS18B20 could be better for precision while sensing the body temperature. For the transmission of sensor values from the wrist band to the receiving kit, other transmission modules such as NRF24L01 or Zigbee can also be used to improve the transmission. For the GPRS Based communication, it would be better to use SIM900 over SIM300.

Security is a major concern these days. While transmitting data from the receiving kit to the server, in this implementation, AES is used with a global key which is not that secure, as if someone gets hold of it, he can access such vital confidential information including locations. For that, any other means of security can be provided such as RSA.

This implementation is very scalable as more sensors can be added to this for collecting other vitals from the patient.

## 5 CONCLUSION

This device is not a medical device to help in treatment but it provides the caregiver to monitor the health condition and also to get alert when in need. Location tracking gives this solution that very required edge so that when a person is in trouble, caregiver can also know where that person is.